\documentclass[10pt]{article}

\usepackage{amsmath}
\usepackage{amssymb}
\usepackage{graphicx}

\title{Limited Holism and Real-Vector-Space Quantum Theory}

\author{Lucien Hardy$^1$ and William K. Wootters$^{1,2,3}$\\
\textit{$^1$Perimeter Institute, 31 Caroline Street North, } \\
\textit{Waterloo, ON N2L 2Y5, Canada}\\
\textit{$^2$Department of Physics, Williams College,} \\
\textit{Williamstown, MA 01267, USA} \\
\textit{$^3$Department of Applied Physics,} \\
\textit{Kigali Institute of Science and Technology,} \\
\textit{BP 3900, Kigali, Rwanda}}

\begin{document}

\maketitle

\begin{abstract}
Quantum theory has the property of ``local tomography'': the state of any composite system can be
reconstructed from the statistics of measurements on the individual components.  In this respect the holism of quantum theory
is limited.  We consider in this paper a class of theories more holistic than quantum theory in that they are constrained
only by ``bilocal tomography'': the state
of any composite system is determined by the statistics of measurements on {\em pairs} of components.
Under a few auxiliary assumptions, we derive certain general features of such theories.  In particular, we show
how the number of state parameters can depend on the number of perfectly distinguishable states.  We also show that real-vector-space
quantum theory, while not locally tomographic, is bilocally tomographic.
\end{abstract}

\section{Introduction}

Standard quantum theory with complex numbers is consistent with a certain local tomography principle: the state of
a composite system can be determined from the statistics of local measurements on the components \cite{Araki, Bergia, Wootters86, Mermin}.  It is not necessary to bring two or more components together and make joint measurements
in order to determine the state.   This property holds in spite of the fact that there are measurements in quantum theory that can
be performed only as joint measurements:
the components must either be physically brought together or be
connected by a quantum channel.   Hence, although quantum theory is, in some sense, a holistic theory, there is a limit to its holism in that states can be accessed locally. The local tomography property has been characterized as capturing the idea that holism can meet reductionism \cite{DAriano}.  It might be thought that, if we drop the assumption of local tomography then there will be no constraining the consequences of holism and any hope of reductionist science goes out the window.  In this paper we show that there are other principles, not as limiting as local tomography, that also
give rise to a kind of limited holism. A general principle one can investigate is $n$-local tomography.

In order to express this principle we introduce the notion of an ``$n$-component measurement.''   Imagine a system partitioned
conceptually into some number of components.  (A given system could be partitioned in many different ways.)  For any
specific choice of this partitioning, we define an $n$-component measurement to be any measurement permitted
by the theory, as long it acts on only $n$ of the components.  In quantum theory, for example, a particular $2$-component measurement might involve an interaction between the two components in question.  Thus a 2-component measurement is more general than a pair of 1-component measurements performed concurrently on two subsystems: in the latter case there can be no interaction between the components.  We now define $n$-local tomography.

\begin{quote}
{\bf $n$-local tomography:} A theory is $n$-locally tomographic if the state of a composite system can be determined from the
statistics of 1-component, 2-component, \dots, and $n$-component measurements.
\end{quote}
This principle becomes a nontrivial constraint on a theory when we have more than $n$ components.  A theory that is $n$-locally tomographic is also, by definition, $(n+1)$-locally tomographic.  The case where $n=1$ is called local tomography and has been investigated extensively.  We will focus on the case where $n=2$ which we will call bilocal tomography.  Standard quantum theory with a complex Hilbert space is locally tomographic.  We will show that quantum theory with a real Hilbert space is bilocally tomographic.  (Real-vector-space quantum theory has been studied in many papers, an early example being Ref.~\cite{Stueckelberg}.)

It is worth emphasizing that we do not assume a privileged decomposition of a given system into subsystems.  In order for
a theory to satisfy the principle of $n$-local tomography, the state of a system must be $n$-locally accessible in the above sense {\em no matter how}
the system is partitioned.  In order to formulate our arguments, though, we will assume that the system has been partitioned into a set of subsystems that are taken as basic for purpose of the discussion.

\section{Basic concepts}

We employ here some elements of the convex probabilities framework that has been developed by various authors \cite{Mackey, Ludwig, DaviesLewis, Araki, Wootters1, FoulisRandall, Gudder, Hardy1, Barrett, CDP, Hardy2}.  The state of a system $A$, in general, can be represented by a list of probabilities associated with a fiducial set of measurement outcomes labeled $k_A\in \Omega_A$,
\begin{equation}
{\bf p}_A = \left( \begin{array}{c} \vdots \\ p_{k_A} \\ \vdots \end{array} \right)  ~~~ k_A \in \Omega_A,
\end{equation}
where this is a minimal list of probabilities that is just sufficient to calculate the probability for a general outcome by means of a linear equation
\begin{equation}  \label{probabilities}
\text{Prob} = {\bf r}_A\cdot {\bf p}_A.
\end{equation}
(By adopting this form for the probability, we are in effect using unnormalized states; the normalization, that is, the
probability of the system being present at all,
will itself be a linear function of the fiducial probabilities.  This choice of representation
turns out to be very convenient for the parameter-counting arguments that follow.)
We define
\begin{equation}
K_A= |\Omega_A|.
\end{equation}
Under the assumption that we can take arbitrary mixtures of preparations, $K_A$ is equal to the the minimum number of probabilities that must be measured to determine the state.  We can regard each value of $k_A$ as corresponding to a filter type measurement, so that $p_{k_A}$ is the probability
that system $A$ will pass through the given filter.

A example is a qubit (say a spin-half particle). Its state can be represented by the list
\begin{equation}
{\bf p}_A = \left( \begin{array}{c} p_{z+} \\ p_{z-} \\ p_{x+} \\ p_{y+} \end{array} \right),
\end{equation}
where $p_{z+}$ is the probability of seeing spin up along the $z$ direction (with similar notation for the other probabilities).  In this case
\begin{equation}
\Omega_A= \{ z+, z-, x+, y+ \}.
\end{equation}
 This list of four probabilities contains the same information as the usual density matrix
\begin{equation}
\widehat{\rho}_A = \left( \begin{array}{cc} p_{z+} & a \\ a^* & p_{z-} \end{array} \right)
\end{equation}
since
\begin{equation}
a = p_{x+} - i p_{y+} - \frac{1-i}{2}(p_{z+} + p_{z-}).
\end{equation}

For a composite system $AB$ the state is given by
\begin{equation}
{\bf p}_{AB} = \left( \begin{array}{c} \vdots \\ p_{k_{AB}} \\ \vdots \end{array} \right)  ~~~ k_{AB} \in \Omega_{AB},
\end{equation}
where we now have $K_{AB}=|\Omega_{AB}|$.

An example of a composite system is a pair of qubits.  Then the state is given by
\begin{equation}
{\bf p}_{AB} = \left( \begin{array}{c} \vdots \\ p_{k_{AB}} \\ \vdots \end{array} \right)~~~ k_{AB} \in
\{ z+, z-, x+, y+ \} \times \{ z+, z-, x+, y+ \}.
\end{equation}
(So a typical member of the list is the joint probability $p_{x+z-}$, which is the probability of the event, ``system $A$ passes through
an $x+$ filter {\em and} system $B$ passes through a $z-$ filter.'')  For this example we have local tomography---we can determine the state of the composite system from local measurements on the parts.

Another integer we will make use of is the maximum number of reliably distinguishable states $N_A$ for a system $A$. This is the maximum number of preparations that can be distinguished in a single shot measurement.  It is equal to the Hilbert space dimension in the case of quantum theory.

\section{Local tomography}

To set ourselves up for a study of bilocal tomography, we will first look at local tomography.
\begin{quote}
{\bf Local tomography:}  A theory is locally tomographic if the state of a composite system can be determined from
the statistics of 1-component measurements.
(Note that the measurements on the separate components may be made concurrently, so that one
can observe correlations among the outcomes.)
\end{quote}
Consider two systems $A$ and $B$.  We can make $K_A$ independent measurements on $A$ and $K_B$ independent measurements on $B$, so that there are $K_AK_B$ distinct combinations.  If we have local tomography we must have covered all possibilities and so we have\footnote{Here
we are making an assumption: that measurements on system $A$ which are operationally equivalent when $A$ is taken alone remain operationally equivalent if $A$ is taken in conjunction with another system, $B$.  In other words, we are assuming that $K_A$ is a non-contextual property of system $A$.  In the foliable joint probabilities framework \cite{Hardy2} such an assumption is unnecessary because system $B$ can be taken as a preparation for $A$ and so any effect of a measurement of $B$ on the value of $K_A$ would already be counted.}
\begin{equation}
K_{AB} \leq K_A K_B.  \label{localtomography}
\end{equation}
We can use this equation iteratively to obtain restrictions on $K$ for more than two systems.  For example, for a system
$ABC$ consisting of three components, we have
$$
K_{ABC} \leq K_{AB}K_C \leq K_AK_BK_C.
$$
Note that it makes no difference if we run this proof grouping $BC$ in the intermediate step.
We can make further progress if we make some additional assumptions.  First we assume
\begin{quote}
{\bf Local independence:} We assume that all the $K_AK_B$ contributions in the above counting are independent---none
are redundant---so that\footnote{Note that the inequality (\ref{localindependence}) does not by itself imply that all
the $K_AK_B$ local contributions are independent.  If local tomography is not valid, then $K_{AB}$ could be large because of
global degrees of freedom, even if many of the $K_AK_B$ locally accessible parameters are redundant.  But Eq.~(\ref{localindependence}) is {\em implied by} local independence. \label{foot}}
\begin{equation}
K_{AB} \ge K_AK_B.  \label{localindependence}
\end{equation}
(In fact this is a very natural assumption that can be obtained from more basic ideas \cite{Hardy2}.)
\end{quote}
When combined with our local tomography
requirement (\ref{localtomography}), local independence implies that
\begin{equation}  \label{independence}
K_{AB} = K_A K_B.
\end{equation}
We can use this equation iteratively to obtain $K$ for more than two systems.  For example
\begin{equation}\label{localiterative}
K_{ABC} = K_{AB}K_C = K_AK_BK_C,
\end{equation}
and again note that it makes no difference if we run this proof grouping $BC$ in the intermediate step.
If we make some further assumptions we can obtain a particular functional form for $K$.  Assume
\begin{quote}
{\bf $N$ - specified:} We assume that, for any system $A$,
\begin{equation}
K_A = K(N_A).
\end{equation}
\end{quote}
\begin{quote}
{\bf Multiplicative:} We assume that the maximum number of distinguishable states satisfies
\begin{equation}
N_{AB}=N_AN_B.  \label{multiplicative}
\end{equation}
\end{quote}
\begin{quote}
{\bf Regular:} We assume that
\begin{equation}
K(N_A+1) > K(N_A).  \label{regular}
\end{equation}
\end{quote}
From these assumptions it follows from the theorem in Appendix 2 that
\begin{equation} \label{K=N^r}
K=N^r,
\end{equation}
where $r$ is a positive integer (similar arguments appear in Refs.~\cite{Wootters86, Hardy1}).

It is easy to see that standard quantum theory (with complex Hilbert space) is locally tomographic.  First we can check the counting.  In an unnormalized density matrix we have $N$ real numbers on the diagonal and $N(N-1)/2$ complex numbers above the diagonal (the numbers below the diagonal are just the complex conjugates of these and therefore contribute no new parameters) and so the total number of independent real parameters is
\begin{equation}
K= N + 2 N(N-1)/2 = N^2.
\end{equation}
This is an example of Eq.~(\ref{K=N^r}) with $r=2$.  To really see that we can do local tomography consider the following projectors on a Hilbert space of dimension $N$:
\begin{equation}  \label{Pn}
\widehat{P}_v = |v\rangle\langle v|  ~~\text{for}~ v=1, \ldots, N;
\end{equation}
\begin{equation}  \label{Pmnx}
\widehat{P}_{uvx} = (|u\rangle + |v\rangle)(\langle u| + \langle v|)~~\text{for}~ v>u;
\end{equation}
\begin{equation}  \label{Pmny}
\widehat{P}_{uvy} = (|u\rangle + i|v\rangle)(\langle u| - i \langle v|)~~\text{for}~ v>u.
\end{equation}
We call a general projector in this set $\widehat{P}_k$ where $k\in \{v, uvx, uvy:  v>u \}$.
There are $N^2$ projectors here and it easy to verify that they are all linearly independent. Hence, they span the space of Hermitian operators acting on a complex Hilbert space of dimension $N$. Therefore the probabilities
\begin{equation}  \label{prob2}
p_k = \text{Trace}(\widehat{P}_k \widehat{\rho})
\end{equation}
are just sufficient to determine the state $\widehat{\rho}$ which provides a proper proof that $K=N^2$ in standard complex Hilbert space quantum theory.  Now, if we have systems $A$ and $B$ comprising system $AB$ then it follows from the nature of the tensor product that the $(N_AN_B)^2$ projectors
\begin{equation}
\widehat{P}_{k_A}^A \otimes \widehat{P}_{k_B}^B
\end{equation}
are all linearly independent and therefore span the space of Hermitian operators on an $N_{AB}=N_AN_B$ dimensional Hilbert space.   Hence the joint probabilities
\begin{equation}
p_{k_Ak_B} = \text{Trace}((\widehat{P}_{k_A}^A \otimes \widehat{P}_{k_B}^B )\widehat{\rho}_{AB})
\end{equation}
are just sufficient to determine the state $\widehat{\rho}_{AB}$ of the composite system $AB$ and therefore we have local tomography (with $K_{AB}=K_A K_B$) for bipartite systems.  By iterating this argument as in Eq.~(\ref{localiterative}), we can extend the conclusion to systems with arbitrarily many components, so that we have local tomography in general \cite{Araki, Bergia, Wootters86}.

%\footnote{!!!Subtle issue needs to go somewhere: We need to assume that the counting is complete as well - in sense that it is possible that measurements on system $A$ which are operationally equivalent when $A$ is taken alone become operationally distinct if $A$ is taken in conjunction with another system, $B$ (call this {\it splitting}). This would effect the counting. However, this possible effect is dealt with in the foliable joint probabilities framework because system $B$ can be taken as a preparation for $A$ and so any such splitting would already be counted.}

\section{Bilocal tomography}

\subsection{Counting contributions}

We now consider bilocal tomography.
\begin{quote}
{\bf Bilocal tomography:} A theory is bilocally tomographic if the state of a composite system can be determined from the
statistics of 1-component and 2-component measurements.
\end{quote}
Consider three systems $A$, $B$, and $C$ for which the numbers of reliably distinguishable states are $N_A$, $N_B$, and $N_C$.
We now count the number of independent parameters that could conceivably be obtained from different types of
bilocal measurements on the combined system $ABC$, in order to derive an upper bound
on $K_{ABC}$ under the assumption of bilocal tomography.  (We use the term ``bilocal measurement'' to refer to
any combination of 1-component and 2-component measurements performed concurrently
on the subsystems.)  First, one can make concurrent 1-component measurements on the three individual
systems.  There are $K_AK_BK_C$ basic measurements of this kind, which together can provide
at most $K_AK_BK_C$ independent parameters.  In addition, one can perform, for example, a joint measurement
(that is, a 2-component measurement) on $AB$ concurrently with an individual measurement on $C$.  Some of the parameters one can obtain by measurements of this form have already been accounted for in the consideration of individual measurements.  But if $K_{AB}$ is greater than $K_AK_B$---that is, if there are joint measurements on $AB$ that are independent of all combinations of local measurements---then one can access additional parameters in this way.  The number of additional parameters provided by such measurements
is at most \hbox{$[K_{AB} - K_AK_B]K_C$}.  Similarly one can measure $AC$ jointly while also measuring $B$, and likewise for $BC$ and $A$.  Altogether then, the maximum number of parameters that could be obtained from all bilocal measurements is
$$
K_A[K_{BC}- K_BK_C] + K_B[K_{AC} - K_AK_C]
+ K_C[K_{AB} - K_AK_B] + K_AK_BK_C
$$
$$
= K_AK_{BC} + K_BK_{AC} + K_CK_{AB} - 2K_AK_BK_C.
$$
By our bilocality axiom, it follows that the value of $K_{ABC}$ is no larger than this sum;
that is, the minimum number of probabilities required to determine the state of $ABC$
cannot exceed what can be obtained bilocally.  So we have
\begin{equation}
K_{ABC} \le K_AK_{BC} + K_BK_{AC} + K_CK_{AB} - 2K_AK_BK_C.  \label{bilocaltomography}
\end{equation}

\subsection{Latent parameters and iteration}\label{latentiteration}

In this section we will show that, if we consider theories satisfying some very natural extra assumptions,
we can obtain some interesting results including that $K=(N^r+N^s)/2$.  The first three extra assumptions are
\begin{quote}
{\bf $N$ - specified:} We assume that, for any system $A$,
\begin{equation}
K_A = K(N_A).  \label{Nspecified}
\end{equation}
\end{quote}
\begin{quote}
{\bf Local independence:} As before, we assume that
\begin{equation}
K_{AB} \ge K_AK_B.  \label{independenceinequality}
\end{equation}
\end{quote}
\begin{quote}
{\bf Bilocal independence:} We assume that all the contributions to $K_{ABC}$ in the above counting are independent,
implying that\footnote{See footnote \ref{foot}.  Eq.~(\ref{bilocalinequality}) does
not by itself imply that the above contributions are independent, but the independence of these
contributions does imply Eq.~(\ref{bilocalinequality}).}
\begin{equation}
K_{ABC} \ge K_AK_{BC} + K_BK_{AC} + K_CK_{AB} - 2K_AK_BK_C.  \label{bilocalinequality}
\end{equation}
\end{quote}
When combined with our assumption of bilocal tomography (\ref{bilocaltomography}), bilocal independence gives us the
equality
\begin{equation}\label{threeparticle}
K_{ABC} = K_AK_{BC} + K_BK_{AC} + K_CK_{AB} - 2K_AK_BK_C.
\end{equation}
From these three assumptions we show in Appendix 1 that for each system $A$ there is a nonnegative real number $L_A$ depending
only on $N_A$,
\begin{equation}
L_A= L(N_A),
\end{equation}
such that
\begin{equation}\label{localpadding}
K_{AB}=K_AK_B + L_A L_B.
\end{equation}
As we will see shortly, under other reasonable assumptions it will turn out that $L_A$ must be an integer.\footnote{Already we know
that $L_AL_B$ must always be an integer, but this fact does not rule out, for example, the possibility that every $L$ is an integer multiple of $\sqrt{2}$.}  In that case it is as if, in addition to the $K_A$ parameters that are locally measurable for system $A$ there are another $L_A$ hidden or latent parameters.  We also derive in
Appendix 1 an analogous equation for $L_{AB}$:
\begin{equation}\label{Lequation}
L_{AB}= K_A L_B + L_A K_B.
\end{equation}

We can use Eqs. (\ref{localpadding}) and (\ref{Lequation}) iteratively to obtain $K$ and $L$ for more than two systems. For example
\begin{equation}\label{bilocaliterative}
K_{ABC} = K_{AB}K_C + L_{AB}L_C = K_AK_BK_C + L_AL_BK_C + K_AL_BL_C + L_AK_BL_C
\end{equation}
It makes no difference which pair of systems we group.
The general form of $K_{AB\dots}$, for an arbitrary number of systems, is such that {\it every possible term with an even number of $L$'s is present}.   Similarly, it is easy to see that the general form of $L_{AB\dots}$, for an arbitrary number of systems, is such that {\it every term with an odd number of $L$'s is present}.

\subsection{Functional form}

Let us now further assume
\begin{quote}
{\bf Multiplicative:} We assume that the maximum number of distinguishable states satisfies
\begin{equation}
N_{AB}=N_AN_B.
\end{equation}
{\bf Regular $+$:} We assume that
\begin{equation}
K(N_A+1) + L(N_A+1) > K(N_A) + L(N_A). \label{sumregular}
\end{equation}
{\bf Regular $-$:} We assume that
\begin{equation}
K(N_A+1) - L(N_A+1) > K(N_A) - L(N_A). \label{differenceregular}
\end{equation}
\end{quote}
Then we can show that
\begin{equation}\label{solution}
K= \frac{1}{2} (N^r +N ^s);  \hspace{3mm} L= \frac{1}{2}(N^r-N^s),
\end{equation}
where $r$ and $s$ are integers satisfying $r \ge s > 0$.  So under these assumptions $L$ must be an integer.
Standard (complex Hilbert space) quantum theory corresponds to the case where $r=s=2$ (so $K=N^2$ and $L=0$).  Real Hilbert space quantum theory corresponds to the case where $r=2$ and $s=1$ (so $K=\frac{1}{2}N(N+1)$ and $L=\frac{1}{2}N(N-1)$).

To obtain Eq.~(\ref{solution}), take the sum and difference of Eqs.~(\ref{localpadding}) and (\ref{Lequation}):
\begin{equation}
K_{AB} + L_{AB} = [K_A + L_A] [K_B + L_B]  \label{summultiplicative}
\end{equation}
and
\begin{equation}
K_{AB} - L_{AB} = [K_A - L_A] [K_B - L_B].  \label{differencemultiplicative}
\end{equation}
That is, both the sum and difference of $K$ and $L$ are multiplicative.  From this, using the above assumptions, we show in Appendix 2 that
\begin{equation}
K(N)\geq L(N)
\end{equation}
$$
K(N) + L(N) = N^r
$$
and
$$
K(N) - L(N) = N^s,
$$
where the integers $r$ and $s$ satisfy $r\ge s >0$.  The last two equations give Eq.~(\ref{solution}) as required.  The point of making the assumptions (\ref{sumregular}) and (\ref{differenceregular}) is to rule out
pathological solutions that involve raising different prime factors to different powers.

\subsection{Proof that real-vector-space quantum theory is \\bilocally tomographic}

We will now prove that quantum theory with real Hilbert space is bilocally tomographic.  First we check the counting.  The density matrix (unnormalized) has $N$ real numbers along the diagonal and $N(N-1)/2$ numbers above the diagonal all of which are real since this is real Hilbert space.  The numbers below the diagonal are equal to their counterparts above the diagonal (since the density matrix must be symmetric).  Hence, there are
\begin{equation}
K= N+ 1 N(N-1)/2 = \frac{1}{2}(N^2+N)
\end{equation}
independent real parameters in the density matrix.  This is an example of Eq.~(\ref{solution}) with $r=2$ and $s=1$.  Hence the counting works.  Eq.~(\ref{solution}) also tells us that the number of latent parameters in this case
should be $L = N(N-1)/2$, and we will see shortly that this is indeed the case.  However, this counting argument is not sufficient to prove that the theory is bilocally tomographic.

To prove bilocal tomography, we consider, for each basic component of our system, the following set of $N^2$ linearly independent Hermitian operators.
\begin{equation} \label{real1}
\widehat{P}_{v} = |v\rangle\langle v| \; \;\hbox{for} \; v = 1, \dots, N;
\end{equation}
\begin{equation}  \label{real2}
\widehat{\sigma}_{uvx} = |u\rangle\langle v| + |v\rangle\langle u| \; \;\hbox{for} \; v > u;
\end{equation}
\begin{equation} \label{imag}
\widehat{\sigma}_{uvy} = -i(|u\rangle\langle v| - |v\rangle\langle u|) \; \;\hbox{for} \; v > u.
\end{equation}
Let us call a general operator listed in any of these three equations $Q_{k}$, where we have $k \in \{v, uvx, uvy : v>u\}$.
The operators defined in Eq.~(\ref{imag}),
being imaginary (and antisymmetric), cannot be used in a linear combination that represents the state of an individual component,
but they will come in handy when we consider larger systems.   Note that there are $N(N-1)/2$ of these
imaginary operators, and $N(N+1)/2$ of the real operators defined in Eqs.~(\ref{real1}) and (\ref{real2}).

Let us label the components of the system $A_1, A_2, \dots, A_m$.  Again, because of the nature of the
tensor product, we know that all operators of the form
\begin{equation} \label{Qproduct}
\widehat{Q}_{k_1\dots k_m}=\widehat{Q}^{A_1}_{k_1} \otimes \widehat{Q}^{A_2}_{k_2} \otimes \cdots \otimes \widehat{Q}^{A_m}_{k_m}
\end{equation}
are linearly independent when regarded as operators on the
complex Hilbert space.  But many of them are imaginary and are therefore not appropriate for representing a
real state.  So we use, out of this set, only those operators for which an {\em even} number of the
indices $k_j$ come from the set $\{uvy: v>u\}$.  The number of operators $\widehat{Q}_{k_1 \dots k_m}$ meeting this condition
is\footnote{This equation follows from an inductive argument.  Let ${\mathcal K}(N_1\ldots N_m)$ be the number of real operators $\widehat{Q}_{k_1 \ldots k_m}$ for the system $A_1
\ldots A_m$, and let ${\mathcal L}(N_1\ldots N_m)= (N_1\ldots N_m)^2 - {\mathcal K}(N_1\ldots N_m)$ be the number of imaginary
operators.   If we assume that ${\mathcal K}(N)=N(N+1)/2$ for some particular value of $m$, it follows that this equation must also be true for $m+1$, since
the functions $K(N)= N(N+1)/2$ and $L(N) = N(N-1)/2$ satisfy Eq.~(\ref{localpadding}).  The equation ${\mathcal K}(N) = N(N+1)/2$ certainly applies to the single system $A_1$.  So by induction, the equation holds for any number of components.}
\begin{equation}  \label{realoperators}
\hbox{(number of real operators)}\, = N(N+1)/2, \hspace{2mm} \hbox{where} \hspace{2mm} N = N_1N_2 \ldots N_m.
\end{equation}
This number is exactly the number of parameters needed to specify a
general real symmetric $N\times N$ matrix.  So the real operators of the form (\ref{Qproduct}) constitute a complete basis for the space of such matrices.

Now, these operators are not all projection operators like the operators in Eqs.~(\ref{Pn}), (\ref{Pmnx}), and (\ref{Pmny}). So they cannot be used to compute probabilities directly as in Eq.~(\ref{prob2}).   Nevertheless, each of these
operators, being real and symmetric, does represent a legitimate observable for the system, and because the operators constitute a complete linearly independent set, the expectation values of these observables are just sufficient to reconstruct the system's state.  Moreover, each of the necessary measurements
can be performed bilocally: each real factor $Q^{A_j}_{k_j}$ represents a
1-component measurement, and any pair of imaginary factors $Q^{A_i}_{k_i} \otimes Q^{A_j}_{k_j}$ represents a 2-component
measurement.  These observations are sufficient to conclude that real-vector-space quantum mechanics is indeed bilocally tomographic.

It is interesting, though, to construct {\em projection operators} with similar properties.  We can do this by modifying each
operator $\widehat{Q}_{k_1\dots k_m}$ via the following steps: (i) For each factor $\widehat{\sigma}^{A_j}_{u_jv_jx}$, replace that factor according to the rule
\begin{equation}
 \widehat{\sigma}^{A_j}_{u_jv_jx} \rightarrow (\widehat{P}^{A_j}_{u_j} + \widehat{P}^{A_j}_{v_j} + \widehat{\sigma}^{A_j}_{u_jv_jx})/2.
\end{equation}
(ii) Partition all the factors of the form $\widehat{\sigma}^{A_j}_{u_jv_jy}$ arbitrarily into pairs, and for each
pair $(A_i, A_j)$, replace $\widehat{\sigma}^{A_i}_{u_iv_iy} \otimes \widehat{\sigma}^{A_j}_{u_jv_jy}$ according
to the rule
\begin{eqnarray}
& \widehat{\sigma}^{A_i}_{u_iv_iy} \otimes  \widehat{\sigma}^{A_j}_{u_jv_jy} \rightarrow  \nonumber \\
& \frac{1}{2}\left[ \left( \widehat{P}^{A_i}_{u_i}  + \widehat{P}^{A_i}_{v_i}  \right) \otimes
 \left( \widehat{P}^{A_j}_{u_j} + \widehat{P}^{A_j}_{v_j} \right) + \widehat{\sigma}^{A_i}_{u_iv_iy} \otimes \widehat{\sigma}^{A_j}_{u_jv_jy} \right].
\end{eqnarray}
The resulting product will then be a projection operator, and it represents a bilocal measurement,
since it is a product of projection operators on one or two components.  Moreover, the linear
independence is not affected by these replacements.  To see this, imagine the operators
$\widehat{Q}_{k_1\dots k_m}$ ordered in a list such that any operator with a larger number of
$\widehat{\sigma}$ factors comes after an operator with fewer $\widehat{\sigma}$ factors.
In the above replacements, each operator $\widehat{Q}_{k_1\dots k_m}$ is replaced by a
linear combination of itself and other $\widehat{Q}_{k_1\dots k_m}$ operators appearing earlier
in this list.  The linear transformation representing these replacements is thus of triangular form, with nonzero
diagonal entries,
and is therefore invertible.  So the new operators inherit linear independence from the original
operators $\widehat{Q}_{k_1\dots k_m}$.
Thus we now have a complete set of linearly independent projection operators for
the real-vector-space theory, each operator
representing a bilocal measurement.

\section{Relaxing the $N_{AB}=N_AN_B$ assumption}

It is interesting to consider relaxing the $N_{AB}=N_AN_B$ assumption.  Physically this corresponds to a situation where \lq\lq extra" distinguishable states come into existence when two systems are combined (which is analogous in some
respects to having $K_{AB}$ not equal to $K_AK_B$ and so is clearly of interest here).  An example would be a die and a key.  The die has $N_A=6$ and the key is like a coin having $N_B=2$.  But we might imagine that the box from which the die is made is actually a lockbox (that is unlocked when the key is proximate) having a coin inside it.  Hence, when systems $A$ and $B$ come together, the total number of distinguishable states of $AB$ is $24$ rather than $12$.

We will not attempt to obtain general results for when $N_{AB}\not=N_AN_B$ but rather restrict to the special case
\begin{equation}
N_{AB} = \alpha N_AN_B
\end{equation}
where $\alpha$ is a positive integer.  Then we simply note that one solution to Eqs.~(\ref{localpadding}) and (\ref{Lequation}) is
\begin{equation}
K= \frac{1}{2}[(\alpha N)^r + (\alpha N)^s ] ; ~~~ L = \frac{1}{2}[(\alpha N)^r - (\alpha N)^s ].
\end{equation}
(To rule out other solutions we would have to make additional assumptions as before.)  Here we require that $r\geq s > 0$ for the integers $r$ and $s$.  The special case where $r=s$ is locally tomographic (it satisfies $K_{AB}=K_AK_B$).

\section{Can we give simple axioms for real quantum theory?}

In \cite{Hardy1} (see also \cite{Hardy2}) a simple set of axioms were given for standard quantum theory (with complex Hilbert space).
The assumption of local tomography (that $K_{AB}=K_AK_B$) was responsible for restricting to the case of complex Hilbert space.  A natural question is can we do the same for quantum theory on real Hilbert space?  We conjecture that the following axioms are sufficient.
\begin{quote}
{\bf Information:} Systems having, or constrained to have, a given information carrying capacity have the same properties.

{\bf Additive:} Information carrying capacity is additive (so $N_{AB}=N_AN_B$).

{\bf Bilocal tomography:} The theory is bilocally tomographic. % but not locally tomographic.

{\bf Continuity:} There exists a continuous reversible transformation between any pair of pure states.

{\bf Simplicity:} For each value of $N$ (starting with $N=1$), $K$ takes the smallest value consistent with the above axioms.
\end{quote}
These are basically the same as the axioms given in \cite{Hardy1} but with bilocal tomography substituting for local tomography.

\section{Discussion}

We have seen that it is indeed possible to relax the reductionism of ordinary quantum mechanics without
moving into the realm of unrestricted holism.  Real-vector-space quantum theory serves as an example of a
theory that goes one step further in the direction of holism, in that it is not locally
tomographic but is bilocally tomographic.  We have also seen that, in a certain sense, the real-vector-space
theory makes maximal use of the freedom allowed by bilocal tomography, in that it satisfies
the principle we have called ``bilocal independence,'' which leads to the equation (\ref{threeparticle}).

It is worth discussing this principle in more depth, as it is more subtle than one might imagine.
Recall that Eq.~(\ref{threeparticle}) was derived
by counting all the bilocally accessible parameters in a three-component system that could conceivably be independent.
The principle then states that all of these parameters are in fact independent.  Meanwhile, the assumption of bilocal tomography asserts
that there are no additional parameters, so that we have the equality expressed in Eq.~(\ref{threeparticle}).  One might suppose that we could have obtained the same equation
(though perhaps with more difficulty) by considering, say, a four-component system or a five-component system.
Conceivably it is possible to do this, but it is not immediately obvious how to do the counting.  In fact, in any bilocally but not locally tomographic theory satisfying Eqs.~(\ref{Nspecified}), (\ref{independenceinequality}), and (\ref{bilocalinequality})---a
theory in which there is by assumption no redundancy in our counting of the bilocally accessible parameters for a
{\em three}-component system---it turns out that
there {\em must} be some redundancy when we apply the same counting method to a four-component system.

A simple example illustrating this point is a collection of four binary systems, $ABCD$, satisfying a theory with $K = N(N+1)/2$ such as real-vector-space quantum theory.  Let us count the number of conceivably independent
parameters obtainable by bilocal measurements.  First, we can make concurrent local measurements on the individual components, giving us
$K_AK_BK_CK_D=3^4 =81$ parameters.  Next, we can make a 2-component measurement together with two
1-component measurements.  Consider, for example, measurements on $AB$ combined with measurements of $C$ and $D$.  The number of conceivably new parameters provided by such measurements is $(K_{AB}-K_AK_B)K_CK_D =
(10-9)(3)(3)= 9$, and we get an equal number of new parameters for each of the other five choices of the 2-1-1 partition, for a total
of 54 parameters from 2-1-1 type measurements.   Finally, we can make a 2-component measurement on each of two pairs.  For each choice of the pairs, the number of new measurements is $(10-9)(10-9) = 1$, and this one measurement could give us one new parameter; there are three ways to choose the pairs, so measurements of this form should provide 3 more parameters.  Thus the total number of independent parameters one might expect to obtain from bilocal measurements on four two-state systems is $81+54+3 = 138$.  However, for the kind of theory we are considering, the number of independent parameters for $N=2^4$ is $K = (16)(17)/2 = 136$.  So two of the parameters that we counted must be redundant.  In the case of real-vector-space quantum theory, it is easy to see why.  If we expand the density matrix in terms of Pauli operators, the new parameter provided by 2-component measurements on the subsystems $AB$ and $CD$ is the coefficient of $(\sigma_y^A \otimes \sigma_y^B)\otimes (\sigma_y^C \otimes \sigma_y^D)$.  But this is the same parameter one could obtain from a measurement on $AC$ and $BD$, or on $AD$ and $BC$.  So those last two measurements provide nothing new.
A similar discrepancy occurs for any other bilocally tomographic theory that satisfies Eqs.~(\ref{Nspecified}), (\ref{independenceinequality}), and (\ref{bilocalinequality}) and that has $L \ne 0$.  In a system with four components, if one treats each measurement on two pairs as if it provided independent parameters, one arrives at a value of
$K$ for the whole system that is too large, that is, larger than the value given by the four-component extension of Eq.~(\ref{bilocaliterative}).  (One adds the term $L_AL_BL_CL_D$ three times rather than just once.)

This example illustrates that it is not a trivial matter to figure out, in a general setting, how many of the parameters that one obtains from a given measurement scheme can be independent.  (Of course one can always answer that question in the
context of a {\em specific} theory.)  The problem becomes even more difficult when one considers, say,
3-local tomography.  In that case it is not obvious whether there exists a natural independence equation analogous to Eq.~(\ref{independence}) in the local case, or Eq.~(\ref{threeparticle}) in the bilocal case.  In Appendix 3 we present one possible approach to this question, but it is not based on counting parameters.  Partly to see how this question of
independence plays out, it would be interesting to
find an example of a theory that is not bilocally tomographic but that does satisfy 3-local tomography.  Such a theory would also shed further light on the general notion of limited holism.

\section*{Acknowledgements}
Research at Perimeter Institute for Theoretical Physics is supported in part by
the Government of Canada through NSERC and by the Province of Ontario
through MRI.

\section*{Appendix 1: The function $L(N)$}

Our aim here is to prove that we can associate with each system a number $L$, which is a function of $N$
satisfying
$K_{AB}= K_AK_B + L_AL_B$ and $L_{AB}= K_AL_B + L_AK_B$.  We assume Eqs.~(\ref{Nspecified}), (\ref{independenceinequality}), and (\ref{threeparticle}).  That is, we assume that $K_A= K(N_A)$, that $K_{AB} \ge K_AK_B$, and that
\begin{equation}
K_{ABC} = K_AK_{BC} + K_BK_{AC} + K_CK_{AB} - 2K_AK_BK_C.  \label{threeparticleagain}
\end{equation}

To begin, consider a collection of four systems
$A$, $B$, $C$, and $D$ for which the numbers of reliably distinguishable states are $N_A$, $N_B$, $N_C$, and $N_D$.  The whole collection must satisfy Eq.~(\ref{threeparticleagain}) when any two of the subsystems are considered as a single entity.  In
particular, for the grouping $\{AD, B, C\}$, we have
$$
K_{ADBC} = K_{AD}K_{BC} + K_BK_{ADC} + K_CK_{ADB} - 2K_{AD}K_BK_C.
$$
And for the grouping $\{A, BD, C\}$, we have
$$
K_{ABDC} = K_AK_{BDC} + K_{BD}K_{AC} + K_CK_{ABD} - 2K_AK_{BD}K_C.
$$
The right-hand sides of these two equations must be equal, since the left-hand sides are; so
we obtain
$$
K_{AD}K_{BC} + K_BK_{ADC} - 2K_{AD}K_BK_C =
$$
$$
K_{BD}K_{AC} + K_AK_{BDC}  - 2K_AK_{BD}K_C.
$$
We now use Eq.~(\ref{threeparticleagain}) to replace the factors $K_{ADC}$ and $K_{BDC}$ with expressions in which each
$K$ factor involves only two of the subsystems $A,B,C,D$.  Making these replacements and collecting like terms, we get
$$
K_{AD}K_{BC} - K_{AD}K_BK_C + K_BK_DK_{AC} =
$$
$$
K_{BD}K_{AC}  - K_AK_{BD}K_C + K_AK_DK_{BC}.
$$
A little further manipulation gives us the following equation.
$$
K_{AD}K_{BC} - K_{AD}K_BK_C - K_AK_DK_{BC} + K_AK_DK_BK_C =
$$
$$
K_{BD}K_{AC} - K_{BD}K_AK_C - K_BK_DK_{AC} + K_BK_DK_AK_C,
$$
in which both sides factor, leading to the simple equation
\begin{equation}
\label{h}
h(N_A, N_D)h(N_B,N_C) = h(N_B,N_D)h(N_A,N_C).
\end{equation}
Here $h$ is defined by $h(N_A,N_B) = K_{AB} - K_AK_B$.  That is, $h(N_A,N_B)$ counts the number
of parameters accessible by a joint measurement on $AB$ beyond what one can
obtain by separate measurements on $A$ and $B$.

What consequence does Eq.~(\ref{h}) have for the form of $K(N)$?  To answer this question,
we consider two cases: either (i) there exists an integer $x$ for which $h(x,x) > 0$, or (ii) $h(x,x)=0$ for all integers $x$.  (Our assumption of local independence guarantees that $h(x,x)$ cannot be negative.)  Consider case (i).  In that case, in Eq.~(\ref{h})
we set both $N_B$ and $N_C$ equal to $x$---here $x$ is a specific integer for which $h(x,x)>0$ (say the smallest such $x$)---and write
$$
h(N_A,N_D) = \left( \frac{h(N_D,x)}{\sqrt{h(x,x)}} \right) \left( \frac{h(N_A,x)}{\sqrt{h(x,x)}} \right).
$$
We can therefore define $L_A$ to be $h(N_A,x)/\sqrt{h(x,x)}$ and conclude that $K_{AD}$ has the form in Eq.~(\ref{localpadding}):
\begin{equation}
K_{AD} = K_AK_D + L_AL_D.
\end{equation}
That is, the number of parameters accessible only by a joint measurement on $AD$ is a product, $L_AL_D$, of factors characteristic of the individual systems $A$ and $D$.

Now consider case (ii): $h(x,x) = 0$ for every integer $x$.  Then Eq.~(\ref{h}), with $N_A = N_C$ and $N_B = N_D$, gives us $h(N_A,N_D)^2 = 0$.  It follows that $h(N_A,N_D) = 0$ for all values of $N_A$ and $N_D$.  This result is still consistent with Eq.~(\ref{localpadding}): we simply set $L_A = 0$ for all $N_A$.  That is, there
are no hidden parameters.  (This case includes ordinary quantum mechanics.)

Note that according to the above definitions $L$ is a function of $N$ (i.e. $L_A = L(N_A)$).  At this stage we cannot
say that $L$ is necessarily an integer.  We know that $L(N) = \sqrt{h(N,N)}$; so each $L(N)$ must be the square root
of an integer.

We now look for an equation analogous to Eq.~(\ref{localpadding}) that determines the value of $L$ for a {\em joint} system.  That is, if we know the values of $K$ and $L$
for each of two systems $A$ and $B$, what do we know about $L_{AB}$?

This question can be answered directly from Eq.~(\ref{threeparticleagain}).  Let us rewrite that equation, substituting for every factor of the form $K_{XY}$ the expression given in Eq.~(\ref{localpadding}).
After a little simplification, the equation becomes
$$
K_{ABC} = K_AK_BK_C + K_AL_BL_C + K_BL_AL_C + K_CL_AL_B.
$$
On the other hand, we can also group the system into $AB$ and $C$ and write
$$
K_{ABC} = K_{AB}K_C + L_{AB}L_C
=[K_AK_B + L_AL_B]K_C + L_{AB}L_C.
$$
Comparing the two equations, we see that
$$
L_C[L_{AB}- K_AL_B - L_AK_B] = 0.
$$
Thus either $L_C$ is zero for every value of $N_C$ (as in quantum mechanics), or
\begin{equation}
\label{Lequation2}
L_{AB} = K_AL_B + L_AK_B.
\end{equation}
In fact Eq.~(\ref{Lequation2}) is true even if $L$ is always zero.

\section*{Appendix 2: Functional forms}

We first want to establish the form of $K(N)$ for the case of local tomography.  We assume
that $K(N_AN_B) = K(N_A)K(N_B)$, which follows from Eqs.~(\ref{independence}) and (\ref{multiplicative}), and that
$K$ is a monotonically strictly increasing function of $N$, which follows from Eq.~(\ref{regular}).

The first assumption, $K(N_AN_B) = K(N_A)K(N_B)$, leaves completely free the value of $K$ for each prime value
of $N$.  For each prime $p$, let $k_p = K(p)$.  It then follows from the multiplicative assumption
that if we write a general value of $N$ in terms of its prime factors---that is, $N = 2^{n_2}3^{n_3}\ldots m_N^{n_{m_N}}$, where $m_N$ is the largest prime factor of $N$---the form of $K$ must be
\begin{equation}
K(N) = k_2^{n_2}k_3^{n_3}\ldots k_{m_N}^{n_{m_N}}.  \label{primes}
\end{equation}
We now use the monotonicity assumption to show that each $k_p$ must have the form $k_p = p^r$, for a fixed
value of $r$ independent of $p$.  We will do the proof by contradiction.

Suppose that for two primes $p$ and $q$, we have $k_p = p^{r_p}$ and $k_q = q^{r_q}$, with $r_p \ne r_q$.
Let $N_A = p^a$ and $N_B = q^b$.  From Eq.~(\ref{primes}) we have that
\begin{equation}
\frac{\ln K_B}{\ln K_A} = \frac{b r_q \ln q}{a r_p \ln p},
\end{equation}
while the corresponding ratio for the $N$'s is
\begin{equation}
\frac{\ln N_B}{\ln N_A} = \frac{b \ln q}{a \ln p}.
\end{equation}
Let us now choose the integers $a$ and $b$ so that the ratio $a/b$ lies strictly between the two real numbers
$\ln q/\ln p$ and $(r_q/r_p)( \ln q/\ln p)$, which are distinct by assumption.  Then the ratios
$\ln K_B/\ln K_A$ and $\ln N_B/\ln N_A$ will lie on opposite sides of the number 1.  That is, a larger value of $N$
will correspond to a smaller value of $K$, contradicting our monotonicity assumption.  It follows that
there is a single value of $r$ such that $k_p = p^r$ for every prime $p$.  Eq.~(\ref{primes}) then tells us that
\begin{equation}
K(N) = N^r.
\end{equation}
The number $r$ must be a non-negative integer to avoid fractional values of $K$.  But strict monotonicity rules out
$r=0$; so $r$ must be a positive integer.

We now turn to the case of bilocal tomography.  Our starting point is very similar to what we started with in the
case of local tomography: multiplicativity and monotonicity.  But now these assumptions apply separately
to the two quantities $K(N) + L(N)$ and $K(N) - L(N)$.  (See Eqs.~(\ref{sumregular}), (\ref{differenceregular}),
(\ref{summultiplicative}), and (\ref{differencemultiplicative}).)  In the case of the difference, before we can use the above argument, we
need to show that $K(N) - L(N) \ge 0$.  To see this, note first that by applying Eq.~(\ref{differencemultiplicative}) to the case $N_A = N_B = 1$, we conclude that $K(1) - L(1)$ is either 0 or 1.
That $K(N) - L(N)$ is never negative then follows from the monotonicity equation, Eq.~(\ref{differenceregular}).

We can now apply the above argument, which shows us that
\begin{equation}
K(N) + L(N) = N^r \hspace{1cm} \hbox{and} \hspace{1cm}K(N) - L(N) = N^s,
\end{equation}
from which, as we have seen, it follows that
\begin{equation}  \label{rs}
K= \frac{1}{2} (N^r +N ^s)  \hspace{1cm}  \hbox{and} \hspace{1cm} L= \frac{1}{2}(N^r-N^s).
\end{equation}
In order that $K$ always be an integer, we must take $r$ and $s$ to be nonnegative integers, and as before,
strict monotonicity requires that each be positive.
Moreover we must have $r \ge s$ since we have defined $L$ to be nonnegative. (In principle, we could have defined $L(N)$ to be
always negative or zero.  That is, in Appendix 1 we could have written $L(N)=-h(N,x)/\sqrt{h(x,x)}$ instead of $L(N)=h(N,x)/\sqrt{h(x,x)}$.  Then in Eq.~(\ref{rs}) we would have $s \ge r$.  Eqs.~(\ref{localpadding}) and (\ref{Lequation}) make clear that this change would have no effect on $K(N)$.  But from an interpretational point of view this would have been an odd choice.)

\section*{Appendix 3: An approach to $n$-local \\independence}

Do the local and bilocal ``independence'' equations, (\ref{independence}) and (\ref{threeparticle}), generalize in a natural way to $n$-local tomography?  Here we present one approach to this question, which will lead
us to a specific equation that one might take to express 3-local independence in a 3-locally tomographic theory.

Let us use
the expression ``locally ideal'' to describe any theory that satisfies Eq.~(\ref{independence}).  (A locally tomographic theory
satisfying the local independence condition is thus locally ideal---the number of conceivably independent parameters accessible by local measurements is exactly equal to the number of parameters needed to specify the global state.)
Similarly, let us use  ``bilocally ideal'' for any theory satisfying Eq.~(\ref{threeparticle}).  We now ask whether one can define a reasonable notion of  ``$n$-local ideality.''  Presumably this condition should be expressed by an equation of the form
\begin{equation}  \label{nlocalcompleteness}
K_{A_1A_2\ldots A_{n+1}} = \sum_P \alpha_P \hbox{(product of the $K$'s of the subsystems given by $P$)},
\end{equation}
where the sum is over all partitions $P$ of the system into subsystems, and $\alpha_P$ is a real number associated
with the partition $P$.  In our earlier examples, each partition corresponded to a particular set of measurements
performed concurrently on the subsystems specified by the partition.  But here we focus more on the mathematical
form of the condition rather than on its interpretation in terms of measurement.

We begin with an alternative set of assumptions leading to Eqs.~(\ref{independence}) and (\ref{threeparticle}).
\begin{quote}
{\bf Permutation invariance:} We assume that the right-hand side of Eq.~(\ref{nlocalcompleteness}) is symmetric under all permutations
of the basic components.
\end{quote}
\begin{quote}
{\bf Triviality:} We assume that there exists a ``trivial system,'' that is, a system such that (i) $K=1$ and (ii) when the system is included as part of a
larger system, it does not affect the value of $K$.
\end{quote}
\begin{quote}
{\bf Novelty:} We assume that the condition for $n$-local ideality does not imply the condition for $m$-local ideality for any
$m < n$.
\end{quote}
Let us show how these assumptions give rise to the condition of local ideality, that is, Eq.~(\ref{independence}).  For $n=1$,
Eq.~(\ref{nlocalcompleteness}) has the form
\begin{equation}
K_{AB} = \alpha K_AK_B.  \label{1localcompleteness}
\end{equation}
We now let component $B$ be the trivial system, so that $K_{AB} = K_A$ and $K_B = 1$.  Then Eq.~(\ref{1localcompleteness}) becomes
\begin{equation}
K_A = \alpha K_A,
\end{equation}
so that $\alpha$ must be unity and we recover Eq.~(\ref{independence}).  (Note that for this simple case we did not
need to assume
either permutation invariance or novelty.)

To derive the condition of {\em bilocal} ideality (Eq.~(\ref{threeparticle})), we start with
\begin{equation}
K_{ABC} = \alpha(K_{AB}K_C + K_{AC}K_B + K_{BC}K_A) +\beta K_AK_BK_C, \label{2localcompleteness}
\end{equation}
which is the most general form allowed by permutation invariance.  We now
let $C$ be the trivial system.  Then Eq.~(\ref{2localcompleteness}) becomes
\begin{equation}
(1-\alpha)K_{AB} = (2\alpha + \beta)K_AK_B.  \label{alphabeta}
\end{equation}
There are now two possibilities: either $\alpha = 1$, or the equation reduces to the form
$K_{AB} = cK_AK_B$ for some constant $c$.  In the latter case $c$ must be 1 (since $B$ could be the
trivial system), and our 2-local ideality condition would reduce to the 1-local ideality condition, contradicting
the novelty assumption.  We conclude that $\alpha = 1$, from which it follows from Eq.~(\ref{alphabeta})
that $\beta = -2$, so that we indeed recover Eq.~(\ref{threeparticle}).

Extending this approach to the case $n=3$, we start with the equation
\begin{eqnarray}
K_{ABCD} = \alpha(K_{ABC}K_D + \cdots) + \beta(K_{AB}K_{CD} + \cdots)\nonumber \\ + \gamma(K_{AB}K_CK_D +\cdots)
+ \delta (K_AK_BK_CK_D),\label{3localcompleteness}
\end{eqnarray}
where the ellipses indicate similar terms with the components permuted. (For example, $\beta$ multiplies
three terms, each being the product of a pair of two-component $K$ values.)  Setting $D$ equal to the trivial system,
we get
\begin{equation}  \label{Dtrivial}
(1-\alpha)K_{ABC} = (\alpha+\beta+\gamma)(K_{AB}K_C + \cdots) + (3\gamma + \delta)K_AK_BK_C.
\end{equation}
And setting both $C$ and $D$ equal to the trivial system, we get
\begin{equation}  \label{CDtrivial}
[(1-\alpha) - (\alpha + \beta + \gamma)]K_{AB}=[2(\alpha + \beta + \gamma) + (3\gamma + \delta)]K_AK_B.
\end{equation}
Now, in order that Eq.~(\ref{CDtrivial}) not reduce to the condition for 1-local ideality, we must
have $(1-\alpha) = (\alpha + \beta + \gamma)$ and $(3\gamma + \delta) = -2(\alpha + \beta + \gamma)$.
But then Eq.~(\ref{Dtrivial}) reduces to the 2-local ideality condition unless $\alpha = 1$.  So by the novelty
assumption, we must have
\begin{eqnarray}
&\alpha = 1 \nonumber \\
&\alpha + \beta + \gamma = 0  \label{constraints1} \\
&3\gamma + \delta = 0. \nonumber
\end{eqnarray}
These equations do not uniquely determine the values of the coefficients.  So the above argument fails to produce
a unique equation expressing 3-local ideality.

We now consider an additional assumption, which is not obviously consistent with the assumptions we
have already made.
\begin{quote}
{\bf Inclusion:} Any theory that is $n$-locally ideal is also $(n+1)$-locally ideal.
\end{quote}
%This assumption is best justified in the case of a theory that is $n$-locally tomographic.  In that case, if the theory is
%also $n$-locally ideal, then a complete set of independent $n$-local measurements presumably provides exactly the right number of parameters for a determination of the state.  In this case $(n+1)$-local tomography provides nothing
%new beyond $n$-local tomography, and nothing new is needed.  So the theory would also be ideal with respect
%to $(n+1)$-local measurements.
The two assumptions ``novelty'' and ``inclusion'' can be naturally
merged into a single assumption that could be called ``strict inclusion'': $n$-local
ideality implies $(n+1)$-local ideality, but the implication does not go in the other direction for any value of $n$.

Note that the inclusion assumption is true for $n=1$: any theory satisfying Eq.~(\ref{independence})
automatically satisfies Eq.~(\ref{threeparticle}).  (So standard quantum theory, which is 1-locally ideal, is
also 2-locally ideal.)  We now insist that the assumption also be true for $n=2$.  That is, we insist that any theory satisfying Eq.~(\ref{threeparticle}) also satisfy Eq.~(\ref{3localcompleteness}).  Our hope is that this requirement will lead us to unique values for the coefficients in Eq.~(\ref{3localcompleteness}).

Consider, then, any 2-locally ideal theory, that is, any theory satisfying
\begin{equation}
K_{ABC}= K_{AB}K_C + K_{AC}K_B + K_{BC}K_A - 2K_AK_BK_C. \label{threeparticleyetagain}
\end{equation}
The equation must still be satisfied if we replace $C$ with a pair $CD$:
\begin{equation}
K_{ABCD} = K_{AB}K_{CD} + K_{ACD}K_B + K_{BCD}K_A - 2K_AK_BK_{CD}.
\end{equation}
We now symmetrize this equation over all permutations (all of the permuted versions of the equation must
also be true), arriving at
\begin{eqnarray}
&K_{ABCD} = (1/2)(K_{ABC}K_D +\cdots)\nonumber \\ &+ (1/3)(K_{AB}K_{CD}+\cdots) - (1/3)(K_{AB}K_CK_D + \cdots).
\end{eqnarray}
In the first of the three terms on the right-hand side, we are free to use Eq.~(\ref{threeparticleyetagain})
again.  Let us make the replacement
\begin{equation}
K_{ABC} \rightarrow \epsilon K_{ABC} + (1-\epsilon)[(K_{AB}K_C + \cdots) - 2 K_AK_BK_C]
\end{equation}
(and similarly for each other triple of components), where $\epsilon$ is a parameter we can freely choose.  We then arrive
at an equation of the form (\ref{3localcompleteness}) with
\begin{eqnarray}
&\alpha = 1/2 + \epsilon  \nonumber \\
&\beta = 1/3 \nonumber \\
&\gamma = -1/3 - 2\epsilon   \label{constraints2}\\
&\delta = 8 \epsilon. \nonumber
\end{eqnarray}
So the theory in question will also be 3-locally ideal if the coefficients in Eq.~(\ref{3localcompleteness})
satsify Eq.~(\ref{constraints2}) for some value of $\epsilon$.   That is, the inclusion assumption will be true
for $n=2$ if we choose such values of the coefficients.

It seems reasonable, then, to adopt this choice as long as Eq.~(\ref{constraints2}) does not conflict with our earlier assumptions.  One can see that Eq.~(\ref{constraints2}) is
in fact consistent with Eq.~(\ref{constraints1}) and that there is exactly one solution: $\epsilon$ must have the value
1/2, and we have $\alpha = 1$, $\beta = 1/3$, $\gamma = -4/3$, and $\delta = 4$.

So we arrive at the following candidate equation that might express the notion of
3-local ideality:
\begin{eqnarray}
K_{ABCD} = (K_{ABC}K_D + \cdots) + (1/3)(K_{AB}K_{CD} + \cdots)\nonumber \\ - (4/3)(K_{AB}K_CK_D +\cdots)
+ 4 (K_AK_BK_CK_D). \label{3}
\end{eqnarray}
That is, in a 3-local tomographic theory, Eq.~(\ref{3}) would presumably express the condition of
3-local independence.

But this equation raises more questions than it answers.  In particular, we have not shown that the inclusion
assumption with $n>1$ is consistent with our other assumptions.  (The above discussion does not settle this question
even for $n=2$.)  Moreover, it would seem impossible to arrive
at fractional coefficients such as $1/3$ simply by counting parameters as we did in Section 4.1.  Still, the approach
we have considered here offers
one way in which one might begin to analyze $n$-local tomographic theories.

\end{document}